\documentclass[10pt]{article}

\usepackage{epsfig}

\newlength{\dinwidth}
\newlength{\dinmargin}
\newcommand{\aver}[1]{\langle \, #1 \, \rangle \mathstrut}
\newcommand{\GeV}{\, \mathrm{GeV}}
\newcommand{\keV}{\, \mathrm{keV}}
\newcommand{\fr}[2]{\left(\frac{#1}{#2}\right)}
\def\eps{\varepsilon}

\begin{document}
\titlepage

\vspace*{4cm}

\begin{center}

{\Large \bf Effects of new long-range interaction: Recombination\\
       of relic Heavy neutrinos and antineutrinos }

\vspace*{1cm} 
\textsc{K.M. Belotsky,  M.Yu. Khlopov, S.V. Legonkov and K.I. Shibaev} \\

\vspace*{0.5cm}
Moscow Engineering Physics Institute,
Moscow, Russia \\[0.5ex]
Center for Cosmoparticle Physics "Cosmion" of
Keldysh Institute of Applied Mathematics, \\
Moscow, Russia \\[0.5ex] 

\end{center}

\vspace*{0.5 cm}

\begin{abstract}
If stable Heavy neutrinos of 4th generation possess their own 
Coulomb-like interaction, recombination of pairs of Heavy neutrinos 
and antineutrinos can play important role 
in their cosmological evolution and lead to observable consequences. In particular, effect of this new interaction in 
the annihilation of neutrino-antineutrino  pairs can account for
$\gamma$-flux observed by EGRET.
\end{abstract}


\section{Introduction}
This work begins systematic study of model of subdominant component 
of dark matter in form of Heavy neutrinos \cite{4N} in the special 
case when this component possesses its own long range interaction.
It is supposed that new interaction is Coulomb-like, being described 
with unbroken $U(1)$-gauge group. 
We call it y-interaction. Its massless gauge boson and charge are 
called y-photon and y-charge, respectively. 
The existence of this interaction as well as of 4th generation (and 
4th neutrinos), possessing it, can naturally
follow from superstring inspired phenomenology \cite{Shibaev}. It is assumed 
according to \cite{4N} that Heavy neutrino ($N$) belongs to new (4th) 
generation of fermions, that the electroweak and strong charges of 
4th generation quarks and leptons are
attributed analogously to other three Standard Model generations and 
that mass of 4th neutrino lies in range about $m=45-80$ GeV. Strict 
conservation of y-charge implies the lightest fermion of 4th 
generation (4th neutrino) to be absolutely stable. Due to y-charge 
4th neutrino cannot have Majorana mass, while its Dirac mass is of 
the order of other fermions of 4th generation. Thus y-interaction 
provides natural basis for the hypothesis of massive stable 4th 
neutrino.
y-charge neutrality implies neutrinos to be accompanied by the same 
amount of antineutrinos. 

In the early Universe effects of new interaction can influence the freezing out of 
y-charged 4th neutrinos. The existence of y-photons implies y-radiation 
background (y-background), interacting with y-charged neutrinos. 

After decoupling from plasma and 
y-background 4th neutrinos and antineutrinos 
due to their long-range y-interaction can form 
bound systems. These bound systems annihilate 
rather quickly. Therefore such "recombination" of 
4th neutrinos and antineutrinos (N-recombination) 
reduces their relic density and can 
lead to observable effects.

In the Galaxy, 4th neutrino pairs, being strongly subdominant component of Dark Matter (DM), 
can enter the clumps, formed by the dominant DM component and the enhancement of 
N-recombination in such clumps can lead to $\gamma$-flux, saturating high energy diffuse 
$\gamma$-background, observed by EGRET.

The existence of sufficiently longliving y-charged quarks of 4th 
generation \cite{4H} would add new element to the evolution of 4th neutrinos 
and the self-consistent picture for cosmological multi-component 
y-plasma requires separate consideration.
Therefore we consider below the evolution and observable effects 
of 4th neutrinos and y-background 
only. 

\section{y-plasma and y-radiation in early Universe}

In the early Universe at the temperature\footnote
{Throughout in the text the system of units $\hbar=c=k=1$, where 
$k$ is the Boltzman constant, is used.} $T > m$ 4th neutrino pairs 
were in equilibrium with cosmological plasma, what lead to 
exponential decrease of their equilibrium concentration at $T<m$. 
When the annihilation rate became less, than the rate of expansion
(at $T_f\approx m/30$) 4th neutrino pairs were frozen out and 
their amount in the co-moving volume did not change essentially in 
the course of subsequent expansion. However, at lower temperatures, 
after decoupling of the frozen out y-charged 4th neutrinos from 
y-background, 
formation of bound systems neutrino-antineutrino (N-recombination) was 
possible. 

Annihilation timescale of systems with the size $a_b$ bound due to 
interaction with the fine-structure constant $\alpha_y$ is given by
\begin{equation}
\tau_{ann}\sim \frac{m^2a_b^3}{\alpha_y^2}
\label{tann}
\end{equation}
and it is quite short. Therefore such recombination is in fact 
equivalent to effective annihilation of neutrino pairs and it should 
reduce the
number density of primordial 4th neutrinos.

Let us first consider the influence of y-interaction on the process 
of freezing out (at $T\sim m/30$) of 4th neutrinos. New interaction 
opens new channel of 4th neutrinos annihilation, which is analogous 
to $2\gamma$-annihilation of $e^+e^-$
\begin{equation}
N\bar{N}\rightarrow yy.
\label{yy-ch}
\end{equation}
Furthermore for slow 4th neutrinos and antineutrinos (with relative 
velocity $v<2\pi\alpha_y$) it leads to the increase 
of cross section of all their annihilation channels, including one 
via intermediate $Z$-boson ($N\bar{N}\rightarrow Z\rightarrow ...$), 
due to Coulomb-like factor of Sakharov enhancement
\begin{equation}
C=\frac{C_0}{1-e^{-C_0}},\;\;\;\; C_0=\frac{2\pi\alpha_y}{v}.
\label{Coul}
\end{equation}

At neutrino mass of interest ($m\sim 50$ GeV) channel 
$N\bar{N}\rightarrow Z\rightarrow ...$ is 
close to resonance ($m_{res}=m_Z/2\approx 46$ GeV). Therefore the 
effect of channel (\ref{yy-ch}), being suppressed near $m_{res}$, 
grows with the deviation of $m$ from $m_{res}$. 

The ratio of cross sections of annihilation channels of (\ref{yy-ch}) 
and through $Z$-boson
in non-relativistic limit is independent of Coulomb factor and given 
by
\begin{eqnarray}
\frac{\sigma_{yy}}{\sigma_{Z}}\approx 0.144 \fr{\alpha_y}{1/30}^2 
\frac{P_Z(m)}{P_Z(50\GeV)},\\
P_Z(m)=\left(1-\frac{m_Z^2}{4m^2}\right)^2+\frac{\Gamma_Z^2m_Z^2}{16m^4},\nonumber
\label{yzratio}
\end{eqnarray}
where $m_Z$ and $\Gamma_Z$ are the mass and the width of $Z$-boson, 
$P_Z(50\GeV)\approx 0.0289$.
Effect of Coulomb factor also does not lead to the principle change, 
since at $T\sim m/30$ 
velocity of 4th neutrinos is still large. 

Number density of 4th neutrinos $n$ can be expressed through its 
ratio $r$ to the entropy density $s$ 
\begin{equation}
n=rs=r\frac{2\pi^2g_{s}}{45}T^3,
\label{rs}
\end{equation}
where $g_{s}$ takes into account all the species of ultrarelativistic 
bosons and fermions and $T$ is the temperature of photons.

On the figure 1 the frozen out number density of 4th neutrinos 
($r=\frac{n}{s}$)  is presented. 

Both effects of channel (\ref{yy-ch}) and Sakharov enhancement induce 
correction of density of 4th neutrinos within the factor 2-10.

\begin{figure}
\begin{center}
\centerline{\epsfxsize=8cm\epsfbox{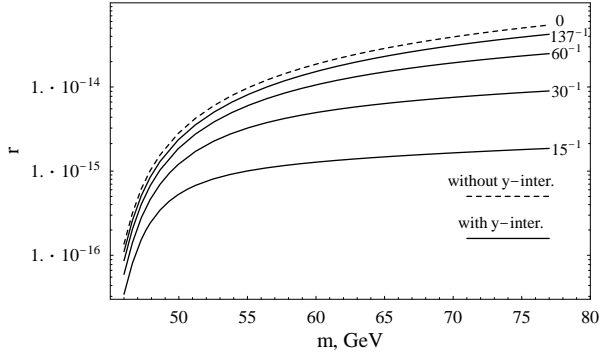}} 
\caption{Number densities of 4th neutrinos frozen out at 
$T\sim m/30$ in units of entropy density 
for the cases without and with y-interaction with different 
magnitudes of its constant.}
\end{center}
\end{figure}

After freezing out the gas of 4th neutrino pairs is for some period 
in thermal equilibrium with ambient ordinary matter 
and y-photon background. 

y-background has direct thermal exchange with 4th neutrino pairs only 
and due to their small frozen out concentration it can not experience 
essential thermal exchange and be in thermal equilibrium
with other matter (ordinary plasma and radiation) at $T<T_f$. 
In the course of successive evolution the number density of y-photons reduces 
relative to that of photons by a factor $\kappa<1$. Due to entropy conservation,
the less species remain in equilibrium with radiation, the smaller is 
the value of $\kappa$. In the period of Big bang nucleosynthesis (at 
$T\sim 1$ MeV) $\kappa\approx 1/6.5$, and y-radiation can not influence 
signficantly the predictions for light element abundance. 

4th neutrinos interact with each other due to long-range 
y-interaction. This interaction with a large, Rutherford-like cross 
section provides 
their equilibrium (Maxwell) distribution, whatever energy losses 
or/and exchanges with other matter components they
experience. 

Before decoupling of 4th neutrinos from y-radiation their
temperature $T_N$ was equal to that of y-photons, i.e. 
$T_N=\kappa^{1/3} T$. After decoupling of 4th neutrinos from 
y-background
(at $T\ll T_{Ny}$) their temperature 
\begin{equation}
T_N= \frac{T^2}{T_{Ny}},\;\;\; T_{Ny}\approx 22\, {\rm keV} 
\fr{m}{50\GeV}^{3/2}\frac{1/30}{\alpha_y}.
\label{TNy}
\end{equation}

The probability for 4th neutrinos and antineutrinos to form their 
bound states grows 
when particles slow down, so at $T\ll T_{Ny}$ the rate of this 
process increases.
For estimation of this effect we will use classical approximation 
following \cite{monopole}. 
$N$ and $\bar{N}$ moving towards each other due to y-attraction 
must experience dipole emission of y-radiation.
If this radiative energy loss exceeds the initial energy of their 
relative motion, they become bound.
The typical length for such energy loss is much less
than the mean distance between 4th neutrinos ({\it a fortiori} than Debye 
radius of N-plasma), 
and the timescale for this energy loss is much less than the 
timescale of N-N and N-y interactions. It provides
such binding to proceed freely, independent of 
other interaction processes.
Cross section of bound system formation in this approach \cite{LLII} 
is obtained to be
\begin{equation}
\sigma_{b}=\pi \rho_b^2(v)=\frac{(4\pi)^{7/5}a_{cl}^2}{v^{14/5}},
\label{sb}
\end{equation}
where $\rho_b$ is the maximal impact parameter at which $N\bar{N}$ 
pair is bound due to y-radiative energy loss,
$v$ is the initial $N\bar{N}$ relative velocity
and $a_{cl}=\alpha_y/m$. 

The rate of binding with cross section having the form 
$\sigma_b=\sigma_0/v^{\beta}$
for particles, distributed by Maxwell, is
\begin{equation}
\Gamma_{rec}=\aver{\sigma_b v}n=
\frac{4\Gamma(2-\frac{\beta}{2})}{2^{\beta}\sqrt{\pi}}\sigma_0\left(\frac{m}{T_N}\right)^{\frac{\beta-1}{2}}rs.
\label{Grec}
\end{equation}

After $N-y$ decoupling  
$\Gamma_{rec}$ exceeds the expansion rate $H$ at
$T$ below
\begin{equation}
T_{rec}\approx 15\,{\rm keV} \fr{m}{50\GeV}^{\frac{5}{16}} 
\fr{\alpha_y}{1/30}^{\frac{11}{8}} \fr{r_{rec}}{1.22\cdot 10^{-15}}^{\frac{5}{4}}
\label{Trec}
\end{equation}
In this expression 
$r_{rec}$ denotes the value of $r(m,\alpha_y)$ before recombination starts, 
being determined by fig.1, and
$r(50\GeV,1/30)=1.22\cdot 10^{-15}$.

The most of created bound systems have typical size less than 
$\rho_b\sim a_{cl}/v^{7/5}\sim a_{cl} (m/T_N)^{7/10}$; 
and their annihilation timescale Eq.(\ref{tann}) turns out to be much less 
than the timescale of their destruction as well as than cosmological 
timescale.

During recombination, 4th neutrino gas will be effectively 
heated, since slow pairs recombine more effectively and
disappear. From the other hand, due to dipole 
emission of $N\bar{N}$ pairs scattered at $\rho > \rho_b$ without binding, 
$N\bar{N}$-gas cools down. Evolution of 4th neutrino pairs with the account 
for the both thermal effects
can be described by the system of equations
\begin{equation}
\left\{
\begin{array}{l}
\frac{dr}{dT}=\aver{\sigma_b v}\frac{r^2s}{HT}\nonumber\\
\frac{d\theta}{dT}=-\frac{2}{3}\frac{\theta}{T_N}\aver{(\frac{3}{2}T_N-E-\frac{1}{3}E_{rel})\sigma_bv}\frac{rs}{HT}.
\end{array}
\right.
\label{dr}
\end{equation}
Here variable $\theta$ shows deviation of neutrino temperature from 
Eq.(\ref{TNy}) due to thermal effects,
$E$ and $E_{rel}$ are the kinetic energy of a single neutrino (or 
antineutrino) and of 
relative motion in $N\bar{N}$ pair, respectively.
\begin{equation}
\aver{(\frac{3}{2}T_N-E-\frac{1}{3}E_{rel})\sigma_bv} 
=\frac{3}{2}\gamma \aver{\sigma_b v}T_N,
\label{esvdip}
\end{equation}
where $\gamma=\frac{5\beta-11}{18}=\frac{1}{6}.$ Note that the terms 
$\aver{(\frac{3}{2}T_N-E)\sigma_b v}$ and $\aver{(-\frac{1}{3}E_{rel})\sigma_b v}$
in Eqs.(\ref{dr},\ref{esvdip}) describe thermal effects due to pair disappearance (heating at $\beta>1$)
and due to dipole emission (cooling), respectively, and the first one prevails ($\gamma>0$). The system 
(\ref{dr}) can be solved by iterations (dividing the first equation 
by the second equation one gets $r(\theta)=r_{rec}\,\theta^{-1/\gamma}$ and 
then subsitutes this value into the system again).
It gives for $r$ at RD stage (at $T>T_{RM}\approx 1$ eV) 
\begin{equation}
r=\frac{r_{rec}}{ \left\{1+F_R(T,T_{rec})\right\}^{1/\bar{\gamma}}}
\label{rRD}
\end{equation}
and at MD stage (at $T<T_{RM}$)
\begin{equation}
r=\frac{r_{rec}}{ \left\{1+F_R(T_{RM},T_{rec})+F_M(T,T_{RM})\right\}^{1/\bar{\gamma}}}.
\label{rMD}
\end{equation}
The functions $F_{R,M}$ have the form
\begin{equation}
F_{R,M}(T,T_0)=r_{rec}\frac{\bar{\gamma}}{\beta_{R,M}}D_{R,M} 
\left(\frac{1}{T^{\beta_{R,M}}}-\frac{1}{T_0^{\beta_{R,M}}}\right).
\label{F}
\end{equation}
Here $\beta_R=\beta-2=\frac{4}{5}$, $\beta_M=\beta-\frac{5}{2}=\frac{3}{10}$, 
$\bar{\gamma}=1+\gamma\frac{\beta-1}{2}=\frac{23}{20}$, 
\begin{equation}
D_R=D_M\sqrt{T_{RM}}=\aver{\sigma_b v}\frac{ (\theta T^2)^{\frac{\beta-1}{2}} s}{H_RT}\approx
1.6\cdot 10^{18}\,{\rm eV}^{\frac{4}{5}}  \fr{m}{50{\GeV}}^{\frac{1}{4}}  \fr{\alpha_y}{1/30}^{\frac{11}{10}}.
\label{D}
\end{equation}
Hubble constants on RD and MD stages are related as 
$$H_R=H_M\sqrt{\frac{T}{T_{RM}}}=\sqrt{\frac{4\pi^3Gg_{\eps}}{45}}T^2,$$
where $g_{\eps}$ takes into account contribution into energy density of all ultrarelativistic species. 
At $T\ll T_0$ the term $1/T_0^{\beta_{R,M}}$ in Eq.(\ref{F}) can be neglected. 
Also $1+F_R$ can be neglected in Eq.(\ref{rMD}). Note that without the account for thermal effects 
($\theta \underline{=}1$) $\bar{\gamma}=1.$

In the period of galaxy formation gas of 4th neutrino pairs becomes strongly nonhomogeneous. 
Since the contribution of 4th neutrinos into the total density is negligible, they cannot play dynamically important role 
and follow the dominant component of dark matter in the process of development of gravitational instability. 
That is why evolution of 4th neutrinos in this period strongly depends on the model of galaxy formation. 
According to CDM model of galaxy formation at $z\sim 10-20$ the distribution of
Dark matter becomes strongly non-homogeneous at small scales. Such structure can not appear in the model of Hot Dark Matter, 
which is however strongly disfavoured by the observational data. 

In the course of development of gravitational instability the fraction of particles, which remain outside the structure of nonhomogeneities
decreases with time. Therefore N-recombination in homogeneous gas of 4th
neutrino turns out to be suppressed, when galaxies began to form at $T=T_{fin}=10T_{mod}$, where $T_{mod}=2.7$ K.
Recombination of 4th neutrinos in Galaxy with the account for small scale
nonhomogeneity (clumpiness) of their distribution is considered in Section 4. 

On Figure 2 the relic densities of 4th neutrinos are shown in units 
of critical density for $\alpha_y=1/30$ and $1/60$. For comparison 
such densities 
without effects of recombination and without y-interaction at all are 
put on Fig.2.

\begin{figure}
\begin{center}
\centerline{\epsfxsize=8cm\epsfbox{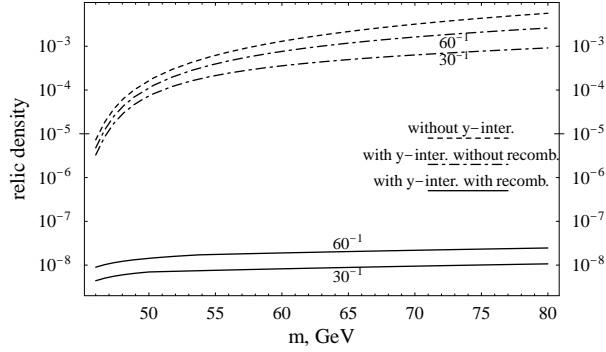}} 
\caption{Relic densities of 4th neutrinos in units of 
critical density for the cases with recombination,
without it and without y-interaction at all. $\alpha_y=1/30,\,1/60$ 
were taken.}
\end{center}
\end{figure}

If (meta-)stable quarks of 4th generation (Q) exist and possess 
y-charge, they represent another component of y-plasma, and joint 
evolution of y-charged N and Q should be studied. However, 
preliminary analysis indicates that there is a range of Q-gas 
parameters, at which presence of Q-component does not influence 
significantly the main features of N-recombination in early Universe.

\section{$\gamma$-emission from recombination of primordial 4th 
neutrinos}

Among the products of N-recombination (and annihilation) on RD stage 
only ordinary (light) neutrinos can survive to the present time. Their maximal modern energy, 
being red-shifted from the period of such early N-recombination, does not exceed few MeV. 
It makes their relatively small flux hardly detectable in neutrino observatories. 

Hadronic and electromagnetic cascades induced by N-recombination interact with the ordinary plasma and radiation. 
The energy, realized in such interaction leads to distortion of CMB spectrum.
If energy release in a unit volume exceeded
$10^{-4}\eps_{\gamma}$ and took place later than at $T\approx 5$ keV 
\cite{Cosm}, it would be observed in CMB spectral measurements.
Total energy realized in the result of recombination at $T<5$ keV is 
given by 
\begin{equation}
\frac {\delta \eps}{\eps_{\gamma}} < \frac{2m 
s(5{\keV}) r(5{\keV})}{\eps_{\gamma}(5{\keV})} \approx
2.5\cdot 10^{-8} \fr{5\keV}{T}^{\frac{7}{23}}
\fr{r_{rec}}{1.22\cdot 10^{-15}}^{\frac{3}{23}}
\fr{50\GeV}{m}^{\frac{5}{23}}\fr{1/30}{\alpha_y}^{\frac{22}{23}}.
\end{equation}
This estimation makes annihilation effects be hardly constrained by the
data on CMB. 

If N-recombination proceeds on MD stage, 
$\gamma$-radiation induced by annihilation in this period will 
contribute into extragalactic $\gamma$-emission 
in the energy range measured by EGRET.

During N-recombination, within interval, when Universe temperature 
falls down on $dT$, $s\cdot dr$ pairs of $N\bar{N}$ in unit volume
annihilate, where $dr$ is given by Eq.(\ref{dr}).

Let $B_Z=\frac{1}{1+\sigma_{yy}/\sigma_Z}$ be fraction of 
annihilation acts going through
the channel $N\bar{N}\!\rightarrow\! Z\!\rightarrow\!...$, being determined 
with the help of Eq.(\ref{yzratio}),
and $dN_{\gamma}(E_0)=f_{\gamma}(E_0)dE_0$ be averaged 
multiplicity 
of created $\gamma$ in this channel within energy interval $E_0 \ldots
E_0+dE_0$. 
Due to redshift modern energy $E \ldots E+dE$ of photons emitted at the 
temperature $T$ (redshift $z$) 
is given by
\begin{equation}
E_0=(z+1)E=\frac{T}{T_{mod}}E,
\label{E0TE}
\end{equation}
Taking into account that number density evolves as $\propto s$, for 
modern number density of $\gamma$ products 
of early $N\bar{N}$ annihilation  we have
\begin{equation}
dn_{\gamma}(E)=B_Z\cdot dN_{\gamma}(E_0)\cdot s(T_{mod})\cdot dr(T)=
B_Zs_{mod}\cdot f_{\gamma}(E_0(T,E))\frac{T}{T_{mod}}dE\cdot r'_TdT,
\label{dng}
\end{equation}
where $r'_T=\frac{dr}{dT}$, $s_{mod}=s(T_{mod})\approx 3\cdot 10^3$ 
cm$^{-3}$.
One can pass from integration $dT$ to $dE_0$. For intensity from 
Eq.(\ref{dng}) we have
\begin{equation}
I_{\gamma}(E)=\frac{c}{4\pi}\frac{dn_{\gamma}}{dE}=
\frac{B_Zs_{mod} c T_{mod}}{4\pi E^2} 
\int_{\frac{T_{fin}}{T_{mod}}E}^{E_{0max}} f_{\gamma}(E_0)\cdot 
r'_T(T(E,E_0))\cdot E_0dE_0.
\label{I}
\end{equation}
Here $E_{0max}$ is the upper limit of annihilation $\gamma$-spectrum, 
$c$ is the light speed. The relationship between $E_0$, $E$, $T$
is given by Eq.(\ref{E0TE}). 

Using results of previous section one obtains intensity of $\gamma$-radiation
from primordial 4th neutrino recombination 
given on Figure 3 for different values of $m$ and $\alpha_y$. 
EGRET data on extragalactic $\gamma$-radiation
are shown for comparison. Spectra $f_{\gamma}(E_0)$ had been obtained with the 
help of code PYTHIA 6.2.

\begin{figure}
\begin{center}
\centerline{\epsfxsize=8cm\epsfbox{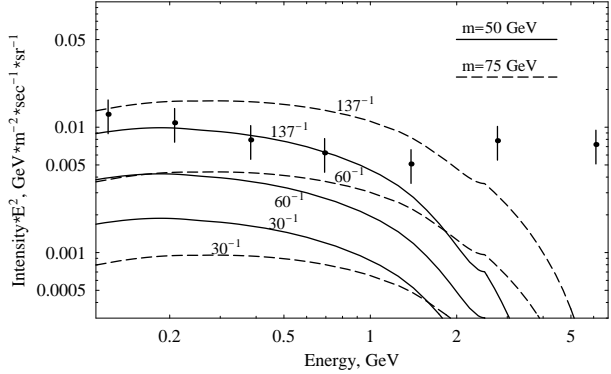}} 
\caption{Gamma fluxes from recombination of 4th neutrinos 
in early Universe in comparison with EGRET data.
$\alpha_y=1/30,\,1/60,\,1/137$ and $m=50,\,75$ GeV were taken.}
\end{center}
\end{figure}

Effect of $T_N$ variation (growth) due to disappearance of slow pairs and to dipole emission was taken into account here.
This effect raises the predicted $\gamma$-flux by factor 2-3.

The growth of $T_N$ reduces $\aver{\sigma_b v}$, what seem to contradict
the mentioned increase in the predicted $\gamma$-flux. 
The fact is that we are interested in annihilation effects produced on the late stages of N-recombination.
Indeed, since the maximal energy of the created 
$\gamma$ is equal to the mass of 4th neutrino, annihilation photons, born in 
period $T < 2000\, {\rm K}< T_{rec}$ only, contribute to the energy range 
of EGRET. To this period the bulk of the primordial 4th neutrinos had been 
annihilated, so the net effect in the interested range is proportional integrally not to 
the initial $r_{rec}$ but to their residual number density, which is in the inverse dependence
on $\aver{\sigma_b v}$.
By the same reason the inverse dependence of predicted $\gamma$-flux on
$\alpha_y$ is explained, being explicitly given by $I_{\gamma}(\alpha_y)\propto 
r'_T(\alpha_y)B_Z(\alpha_y)\propto\alpha_y^{-\frac{22}{23}}r_{rec}^{\frac{3}{23}}(\alpha_y)B_Z(\alpha_y)$. 

Comparison with EGRET data restricts the allowed values 
of $\alpha_y$ as $\alpha_y > 1/137$ for $m=50$ GeV.
This constraint becomes slightly more severe for larger $m$.

Note that the products of N-recombination at $T<2000$ K can provide early ionisation of neutral matter gas, 
which is favoured by WMAP data.

\section{Effects of y-charged 4th neutrinos in Galaxy}

Due to their negligible contribution into the cosmological density (see Fig.2),
4th neutrino cannot play any significant role in the dynamics of galaxy formation. 
Their distribution in Galaxy will be governed by the dominant DM component, which we assume to be CDM, 
for definiteness. In the gravitational potential of Galaxy 4th neutrinos acquire large virial velocities $u \sim 200$ km/s. 
It strongly reduces the rate of their recombination so that the corresponding timescale highly exceeds 
the age of the Universe $t_{mod}$.
However, the typical lifetime of created bound systems in Galaxy is 
much less than $t_{mod}$ and
slow N-recombination can produce $\gamma$-flux, accessible to observations.

Assume that
4th neutrinos are distributed in Galaxy according to isothermal 
halo model with density profile
\begin{equation}
\rho(R)=\rho_{N\,loc}\frac{(1\,{\rm kpc})^2+(8.5\,{\rm 
kpc})^2}{(1\,{\rm kpc})^2+R^2},
\label{rho}
\end{equation}
where $R$ is the distance to the galactic center, 
$\rho_{N\,loc}=\xi\cdot \,0.3$ GeV/cm$^3$. 
The distribution (\ref{rho}) is determined by the dominant CDM
component and the 4th neutrino density is taken to follow it,
being proportional to the ratio $\xi$ of 4th neutrino and dominant CDM densities.
Velocity distribution is assumed to be Maxwellian with the maximal likelihood value 
$u=220$ km/s.
It turns out that the $\gamma$-flux, predicted for various halo models is rather weakly sensitive to their choice, 
varying within a factor 1-2.

The account for 
small scale inhomogeneity (clumps), 
of dominating Cold Dark Matter (CDM) strongly enhances the predicted $\gamma$-flux from N-recombination. 
The possibility of such a 
small scale structure and its survival to the present time is determined by primordial perturbation spectrum 
and on dynamics of galaxy formation \cite{Berez}.
We will take into account by parameter $\eta$
the enhancement of N-recombination (annihilation) rate 
due to clumpiness. In the case of 4th neutrinos such enhancement can be much larger, than for non-interacting DM particles, 
since 4th neutrino annihilation in clumps goes through N-recombination, 
which is enhanced not only due to increase of concentration, 
but also (and dominantly) owing to much smaller virial velocities of 4th neutrinos inside the clump, 
as compared with their averaged density and velocity distributions in the Galaxy.

The following effects are important in the estimation of parameter $\eta$.

Since clumps are formed by the dominating CDM particles, the minimal mass of such clumps is determined 
by the ultraviolet cut in the spectrum of their density fluctuations, which is determined for weakly interacting particles 
by the scale of their free streaming. As in \cite{4N} we'll use the results of \cite{Berez} in  the description of such CDM clumps. 
The existence of the period of thermal equilibrium with ordinary matter and y-radiation makes gas of 4th neutrino hotter, 
than ordinary CDM gas. It prevents 4th neutrinos from entering the smallest CDM clumps and there exists the minimal 
mass of CDM clump,
in which the presence of 4th neutrinos is not suppressed. This critical minimal mass is estimated to be equal to
\begin{equation}
M_{N\,min}\approx 
10^{-2}M_{\odot}\fr{50\GeV}{m}^{15/4}\fr{\alpha_y}{1/30}^{3/2},
\label{MNmin}
\end{equation}
where $M_{\odot}$ is the solar mass. 
Following \cite{4N} we can suppose that 4th neutrino density distribution 
at scales larger than $M_{N\,min}$ is proportional to that of CDM, so 
their relative contribution is 
$\xi=\Omega_{N}/\Omega_{CDM}$, where $\Omega_{N}$ is given by Fig.2, 
$\Omega_{CDM}=0.3$. Radius of clump with
$M=M_{N\,min}$ is equal to
\begin{equation}
R\approx 3\cdot 10^{17}\, {\rm 
cm}\fr{50\GeV}{m}^{5/4}\fr{\alpha_y}{1/30}^{1/2}.
\end{equation}
Typical velocity of particles inside the clump with $M=M_{N\,min}$ is 
determined by
\begin{equation}
u_{cl}=\sqrt{\frac{GM_{N\,min}}{R}}\approx 19\,{\rm m/s}.
\label{ucl}
\end{equation}
We will assume that this velocity parameter is constant over all the volume of clump.
Here we do not consider the possibility for 4th neutrinos to enter the 
clumps of mass less than $M_{N\,min}$,
which was considered in \cite{4N}.

Clumps of mass around the minimal one give the main contribution into 
the annihilation enhancement effect.
The shape of the number density distribution $n$ inside the clump 
determines the effect of annihilation by the factor
\begin{equation}
S= \frac{V}{N^2}\int n^2dV,
\label{Sint}
\end{equation}
where  $V=\frac{4}{3}\pi R^3$, $N=\int ndV$ is the total amount of 
given particles inside the clump; 
$\bar{n}=\frac{N}{V}$ gives the mean number density. 

In case of 4th neutrinos, annihilation rate strongly increases due to 
its velocity dependence so their number in clump
can experience essential change.
In our study of evolution of 4th neutrinos in the clump we will take 
into account
effects of heating due to slow pairs disappearance and of dissipation 
due to dipole emission (the latter turns to be less important, than
the former). Parameter Eq.(\ref{ucl}) is connected with the initial 
temperature of N: $T_{N0}=\frac{mu_{cl}^2}{2}$.
Any decrease of the total number of 4th neutrinos in the clump does not 
affect its gravitational potential
and mass, being determined by dominant DM component. We will assume 
that the shape of distribution of 4th neutrinos inside the clump
does not alter with time, and that the value of the factor (\ref{Sint}) for 4th neutrinos can be taken the same as for non-interacting DM, 
which was estimated in 
\cite{Berez} to be $S \approx 5$.
Then for total amount of 4th neutrinos inside the clump one has
\begin{equation}
\left\{
\begin{array}{l}
\frac{dN}{dt}=-\aver{\sigma_b v}N^2 \frac{S}{V}\nonumber\\
\frac{3}{2}\frac{dT_N}{dt}=\aver{(\frac{3}{2}T_N-E-\frac{1}{3}E_{rel})\sigma_bv}N 
\frac{S}{V}.
\end{array}
\right.
\label{syscl}
\end{equation}
Solution of this system is analogous to that of 
Eqs.(\ref{dr}) and yields
\begin{eqnarray}
N=N_0\fr{T_{N0}}{T_N}^{\frac{1}{\gamma}},\;\; 
T_{N}=T_{N0}\left\{1+\frac{t-t_0}{\tau}\right\}^{\frac{\gamma}{\bar{\gamma}}},\nonumber\\
\dot{N}_{ann}=\frac{\dot{N}_{ann0}}{\left\{1+\frac{t-t_0}{\tau}\right\}^{1+\frac{1}{\bar{\gamma}}}},\;\;
\tau=\frac{1}{\bar{\gamma}\aver{\sigma_b v}_0\bar{n}_0S}.
\label{clpar}
\end{eqnarray}
Here $\dot{N}_{ann}$ means the total annihilation rate given by first 
equation of system (\ref{syscl}). Index "$0$" corresponds 
to the initial moment $T_{N}=T_{N0}$. The mean number density of 4th 
neutrinos in the clump and characteristic time of
recombination in it $\tau$ are estimated as
\begin{equation}
\begin{array}{l}
\bar{n}_0\approx 1.8\cdot 10^{-8} {\rm cm^{-3}} 
\frac{\xi}{2.3\!\cdot\! 10^{-8}} \frac{50\GeV}{m},\nonumber\\
\tau \approx 0.20\, {\rm Gyr}\, \frac{2.3\!\cdot\! 10^{-8}}{\xi} 
\fr{m}{50\GeV}^{\frac{3}{4}}\fr{1/30}{\alpha_y}^{\frac{11}{10}},
\end{array}
\end{equation}
where $\xi(m\!=\!50\GeV,\alpha_y\!=\!1/30)=2.3\cdot 10^{-8}$ was used; 
note that $\xi(50, 1/137)=1.1\cdot 10^{-7}$.
Neglecting the dependence of $r$ on $r_{rec}$, one obtains $\xi\propto 
m^{18/23}\alpha_y^{-22/23}$, so $\tau$ is almost independent 
of $m$ and $\alpha_y$.

The obtained value of $\tau$ exceeds the clump dynamical timescale 
$\sim R/u_{cl}\approx 0.005$ Gyr what provides the shape
of 4th neutrinos distribution to follow the 
"equilibrium" form inside the clump. 
However, the temperature $T_N$ during $t-t_0\sim 10$ Gyr becomes 2 times larger. 
It implies that for the "equilibrium" form of N distribution
the effect of N evaporation can be important. It can lead to the 
slowing down of annihilation rate in the clump as compared with the above 
estimation.

Enhancement factor $\eta$ for 4th neutrinos, corresponding to the present moment may be estimated as 
\begin{equation}
\eta=\eta_0 \fr{u}{u_{cl}}^{9/5} \fr{\tau}{t-t_0}^{1+\frac{1}{\bar{\gamma}}}
<3\cdot 10^4 \fr{m}{50\GeV}^{9/4}\fr{1/30}{\alpha_y}^{9/10}.
\label{eta}
\end{equation}
Here the factor $\eta_0\propto 
\bar{n}_{cl}^2/\bar{n}_{Gal}^2$ (i.e. proportional to the ratio of averaged squared number
densities of DM particles inside the clumps and steeply
distributed in Galaxy) denotes the enhancement of annihilation of non-interacting DM in clumps. 
From figures 3-5 of \cite{Berez} one obtains its value in the range $1 \le \eta_{0} \le 5$ for $M_{min}=M_{N\,min}$.

The estimation (\ref{eta}) does not take into account the mass spectrum of clumps. The latter is predicted to start from the
minimal value $M=M_{min}$ and to fall down with the increase 
of $M$.  The account for this mass distribution
of clumps will slightly decrease the value of $\eta$. 
Indeed, in Eq.(\ref{eta}) the $u_{cl}$
dependence on $M$ gives additional suppression $\propto M^{-3/5}$.

Active annihilation of 4th neutrinos inside the clumps after they separate from the cosmological expansion 
at $z\sim 10-20$ contributes into the intergalactic $\gamma$-background. However, in the period of clump formation,
when the density inside the clumps is of the same order as the average density ($\delta \rho/\rho \sim 1$), 
this contribution cannot exceed the effect of N-recombination by homogeneously distributed 4th neutrinos. 
During the successive periods of galaxy formation the amount of neutrinos inside the clumps decreases due to their annihilation, 
while homogeneously distributed neutrinos enter the inhomogeneities and their fraction reduces with time. It leads to the suppression
of extra-galactic contribution into the high energy part of gamma background as compared with 
the local effect of N-recombination in clumps of the halo of our Galaxy.

Intensity of the modern $\gamma$ radiation from annihilation of 4th neutrinos in the direction $\vec{l}$ 
to halo of Galaxy is determined by
\begin{equation}
I(E)=\eta\cdot \frac{B_Z\aver{\sigma_b 
v}}{4\pi}f_{\gamma}(E)\int_0^{\infty}\fr{\rho}{m}^2dl.
\end{equation}
For $m=50$ GeV, $\alpha_y=1/137$ with factor $\eta \approx 200$ one 
obtains $\gamma$-flux
shown on Figure 4.

\begin{figure}
\begin{center}
\centerline{\epsfxsize=8cm\epsfbox{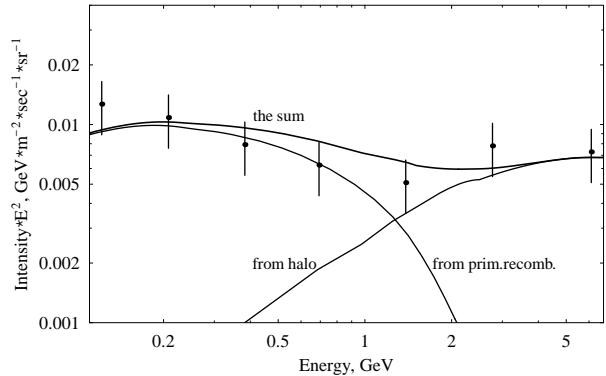}} 
\caption{$\gamma$-fluxes from recombination of 4th 
neutrinos in early Universe and in Galaxy with enhancement factor 200
in comparison with EGRET data for $\alpha_y=1/137$ and $m=50$ GeV.}
\end{center}
\end{figure}

To saturate the high energy part of 
EGRET data at larger values of $\alpha_y$, a slightly larger factor $\eta$ is required. For instance, 
if $\alpha_y=1/30$ (at $m=50$ GeV), $\eta \approx 240$.
Such a weak sensitivity of the predicted galactic $\gamma$-flux to the value of $\alpha_y$ follows from the fact 
that this flux depends on $\alpha_y$ virtually only through such dependence of $B_Z$. The latter is weak for the values of $m$
close Z-boson resonance. It strongly differs from the case of primordial recombination, 
where in addition neutrino velocity depends on $\alpha_y$ through $T_{Ny}$.
That is why a slight change of the factor $\eta$ can easily compensate the decrease of $B_Z(\alpha_y)$ for larger $\alpha_y$.

\section{Discussion}
The existence of new long range interaction would lead to new forms of 
matter, such as y-plasma and to a set of new phenomena,
in which N-recombination, considered in the present paper
plays important role.

In the considered case of y-charged 4th neutrinos N-recombination
suppresses strongly the pre-galactic density of this subdominant
component of Dark Matter, but appears possible to
explain all the data on 
extragalactic $\gamma$-emission observed by EGRET 
by the effect of N-recombination for $m\approx 50$ GeV and 
$\alpha_y\approx 1/100$.

It should be noted that in the framework of Grand Unification 
(GUT) models the value of $\alpha_y$
can hardly be less than $\alpha_{em}=1/137$. Moreover,
such a small value of $\alpha_y$
implies rather specific requirements
for GUT model, which embeds y-interaction.
Indeed, y-interaction, being possessed by 4th 
generation fermions only, should be unified with other interactions 
at large energy scale. Therefore one can expect that it should
have constant at low energies greater than electromagnetic one, 
since the last is reduced from the unified (GUT) constant
due to screening by virtual particles of all the four generations.

The future theoretical model, which will embed the considered here new long range U(1) interaction, 
should satisfy a set of necessary conditions, such as anomaly freedom of y-interaction. 
The y-charge assignment $e_{y}$ of 4th generation particles (N, E, U, D) like 
$e_{yN}=e_{yE}=-e_{yU}/3=-e_{yD}/3$ would avoid triangle anomalies
within Standard Model field content extended to four generations.
However, triangle anomaly problem, as well as the possible effects of $y$-$\gamma$, $y$-$Z$ mixings \cite{y} 
should be considered within a unified framework. To avoid theoretical inconsistencies such framework 
can involve possible extension of particle content 
(SUSY partners, exotics of $E_6$ superstring inspired GUT model, etc.), what 
requires separate study.

Taking into account these theoretical problems, the present analysis has no aim to give any final conclusion. 
It just opens the room for the systematic search for WIMPs, possessing 
new long range forces. One can apply to these searches the methods, 
developed in this work. One can hardly imagine the impact of the 
existence of such particles. Reminding Zeldovich, we can only repeat 
his words: "Though the probability for these phenomena is small, their 
expectation value is great!"

\section{Acknowledgement}

M.Kh. is grateful to the CRTBT-CNRS, Grenoble, France for 
hospitality.

\end{document}